\documentclass[aps,prc,twocolumn,superscriptaddress,showpacs,showkeys]{revtex4-1}
\usepackage{graphicx}
\begin{document}
	\title{Monte-Carlo sensitivity study for sterile neutrino search with $^{144}$Ce~--~$^{144}$Pr source and liquid scintillation detectors of various geometries}
	
	

	\author{A.~V.~Derbin}
	
	\affiliation{Petersburg Nuclear Physics Institute, Gatchina 188350, Russia \\ National Research Center "Kurchatov Institute"}
	
	\author{ I.~S.~Drachnev}
	
	\affiliation{Petersburg Nuclear Physics Institute, Gatchina 188350, Russia \\ National Research Center "Kurchatov Institute"}
	
	\author{I.~S.~Lomskaya}
	
	\affiliation{Petersburg Nuclear Physics Institute, Gatchina 188350, Russia \\ National Research Center "Kurchatov Institute"}
	
	\author{ V.~N.~Muratova}
	
	\affiliation{Petersburg Nuclear Physics Institute, Gatchina 188350, Russia \\ National Research Center "Kurchatov Institute"}
	
	\author{N.~V.~Pilipenko }
	
	\affiliation{Petersburg Nuclear Physics Institute, Gatchina 188350, Russia \\ National Research Center "Kurchatov Institute"}
	
	\author{D.~A.~Semenov}
	
	\affiliation{Petersburg Nuclear Physics Institute, Gatchina 188350, Russia \\ National Research Center "Kurchatov Institute"}
	
	\author{E.~V.~Unzhakov}
	
	\affiliation{Petersburg Nuclear Physics Institute, Gatchina 188350, Russia \\ National Research Center "Kurchatov Institute"}
	
	\begin{abstract}
		Expected energy spectra calculations for large volume liquid scintillation detectors to inverse $\beta$-decay for antineutrinos produced by $^{144}$Ce -- $^{144}$Pr artificial source have been performed.
		The calculations were carried out through Monte-Carlo method within GEANT4.10 framework and were purposed to search for neutrino oscillation to sterile eigenstate with mass about 1 eV.
		The analysis of relative sensitivity to oscillation parameters for different detector shapes has been performed.
	\end{abstract}
	
	
	\maketitle
	
	\section{Introduction}

The solar neutrinos were first registered by Homestake River experiment~\cite{homestake} in the beginning of 1970s.
When the first results had been obtained, it appeared that the solar neutrino flux is significantly smaller than it was expected from the solar model.
This fact has initiated a new challenge for the solution of so-called Solar Neutrino Problem that was later solved with the theory of neutrino oscillations~\cite{Pontecorvo1958, Maki1962, gribov1969, Wol1978, MS1985}.
Experimental verification of this theory involved reactor~\cite{reactor_review}, radiochemical~\cite{SAGE_results,GALLEX_results} and water \v Cherenkov~\cite{SK1999, SNO} neutrino experiments that had discovered neutrino oscillation to non-electron flavors.
At the same time, it was observed that at small distances there is a deficit of electron neutrinos~\cite{SAGE, GALLEX, r_anomaly}, although statistical significance of this deficit is relatively small.
The existence of another neutrino eigenstate with mass around $1$~eV could possibly explain this observation.

In this case the neutrino deficit at small values $L/E_\nu$ would have been explained by neutrino oscillation to this state.
Such state has to be sterile since the number of neutrino flavors is fixed to 3 according to experiments on Z-boson decay width~\cite{width}.
A possibility of new Beyond Standard Model (BSM) physics existence in the neutrino sector is a very strong motivation for sterile neutrino search.
A reactor experiment would be an essential mean of performing this search, although the involvement of a nuclear reactor would introduce significant systematics, which are difficult to control.

An experiment using a radiochemical neutrino source would produce more reliable results despite much smaller achievable statistics, thus it makes sense to perform a source-based sterile neutrino search in parallel with reactor experiments.
Here, we study a possibility of sterile neutrino search with the antineutrino source of $^{144}$Ce~--~$^{144}$Pr, which is quite promising due to high endpoint energy of $^{144}$Pr ground state beta-transition and possibility of strong background suppression due to inverse beta-decay(IBD) reaction signature in liquid scintillators.   
Unfortunately, two excellent proposed and advanced projects using large volume liquid scintillator detectors BOREXINO~\cite{Bx_SOX}  and KamLAND~\cite{Celand} have not been implemented.    

\section{Options for source-based sterile neutrino search}

The most frequently used antineutrino detection mechanism is the inverse beta decay reaction on hydrogen nuclei.
It allowed F.Reynes and C.Cowan to discover neutrinos in the first place back in 1956~\cite{1956Natur.178..446R}, and it still remains one of the most advantageous reactions for antineutrino detection nowadays.

In this process the antineutrino is captured on a proton with consequent production of a neutron and a positron that gets the most of the initial kinetic energy due to its much smaller mass.
The main advantage of this reaction is production of a neutron that is captured after thermalization that takes around $200$~$\mu$s in absence of neutron capturer (e.g. Gadolinium).
The neutron capture on a nucleus results in a deexcitation gamma-quantum emission that could be detected.
Thus, one observes a prompt signal from the positron detection and a delayed signal from the gamma-emission having a clear signature of an antineutrino detection. Such feature of this reaction allows to strongly decrease the background In a large number of cases an anti-neutrino experiment using IBD manages to operate in zero-background mode.

Since the reaction of $^{144}$Pr beta-decay has the endpoint energy of 3~MeV, the maximum positron energy is $\approx1.2$~MeV the total energy deposit in the detector would spread from $2 \times m_0 c^2$ to $\approx2.2$~MeV.
This limits the detector choice options to radiochemical detectors and organic scintillators as water \v Cerenkov detectors can not efficiently detect such low-energy particles.
Organic scintillators have the advantage of online DAQ and energy and position reconstruction that make the scintillator option choice quite essential. 

\begin{figure}[ht]
	\begin{center}	
		\includegraphics[bb=60 20 570 295,width=7cm,height=3.5cm]{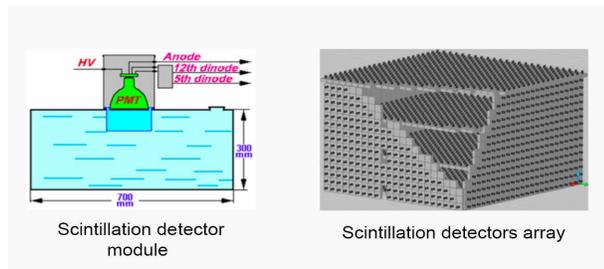}
		\caption{Principal scheme of Baksan Neutrino Observatory detector.
			The right picture shows the general construction of the detector array, consisting of 3180 liquid scintillation (LS) detectors with 330 t total mass.
			The left picture shows the principal scheme of the LS detector, equipped with a single PMT-49 \cite{BNO2}.(figures are taken from Baksan Laboratory website \cite{BNO2}) }
		\label{Fig:BNO}
	\end{center}
\end{figure}

In this work we have performed a study of possibilities to search for neutrino oscillation to sterile eigenstate with $^{144}$Ce~--~$^{144}$Pr antineutrino source and various detector geometries in order to establish the optimal one.
As for the purpose of fair comparison we have considered the same size of the laboratory hall and thus the same linear dimensions of the setup of $6$~m. 
We have performed a test of sensitivity to neutrino oscillation to sterile eigenstate for the following detector geometries: cylinder, sphere and a cube with linear sizes of 6 m as well as an existent detector implementation, Baksan Neutrino Observatory (BNO)~\cite{Kuzminov2012}. 
BNO is a scintillator detector located in a gallery with volume of 12000 m$^3$ in 550 m from the entrance with passive shielding thickness of 850 m.w.e.
It is shaped as a 4-stage building with basement of 16.7 x 16.7 $\rm{m}^2$ and height of 11.1 m., see fig.\ref{Fig:BNO}.
The setup is constructed of low-radioactivity concrete and is fitted with 3180 small white-spirit scintillator tanks with total scintillator mass of 330 tons.
This setup has the advantage of $4\pi$ geometry since the source could be located in the center of the detector but it has relatively small sensitive volume thickness along antineutrino trajectory.
The actual setup in the current state would suffer from relatively high threshold ( $8$~MeV) and inability to record individual detector signals, but in this work we neglect this fact since our interest is to test the maximum achievable sensitivity with any detector upgrade possible (up to zero energy threshold for IBD).

Since various liquid scintillators have practically same hydrogen density in order to make a test on the detector geometry only, it was decided to make the evaluation for the same scintillating liquid, the one named "White-spirit" also known as solvent naphtha, petroleum spirits, Varsol.
This is organic mixture of aliphatic, open-chain or alicyclic C$_7$ to C$_{12}$ hydrocarbons. This scintillator is already used in the BNO detector. 

\section{Simulation and sensitivity study}
 
The sensitivity test of the scintillator detectors has been performed via Monte-Carlo method simulation within a custom-developed software based on Geant4.10.04 framework.
We were considering the IBD cross section according to~\cite{Strumia}. 
The calculation of electron antineutrino spectrum taking into account the corrections for beta-electron interactions with nucleus and atomic shell were based on work \cite{Huber}.
New precision measurements of  $^{144}$Ce--$^{144}$Pr beta-spectrum are taken now in order to determine the antineutrino energy spectrum \cite{NIM_PNPI}. 
Hydrogen atom density for the scintillator used was taken as 6.77 $\times 10^{22} \ \rm{cm}^{-3}$.
We made a simplification of zero source size that is not very important for the result due to spatial resolution of the detectors.
Moreover, we neglected the effect of gamma-quantum escape from the scintillator, that could play some minor role in the case of BNO geometry.
The simulation output is the antineutrino spectrum S(E,R), where E is prompt event energy deposit and R is the distance from the source location. 
The spacial and energy resolutions are taken as 13 cm and 5\% at 1 MeV of energy deposit and are scaled as the square root of $E_{promt} /1 \rm{MeV}$.
In case of BNO detector array we assume only knowledge of the exact tank where antineutrino is registered and no spatial resolution within its volume.

The sensitivity estimate of an experiment was performed for the source statistics corresponding to 1.5 years of exposure with a $^{144}$Ce--$^{144}$Pr source with activity of 100 kCi and zero-background operation regime. 
The source was located in 2 m from the detector sensitive volume assuming presence of 2 m thick layer of water that plays a role of passive shielding for external neutrons and gamma-radiation.
The BNO detector is tested with 50 kCi source that is currently available.

\begin{figure}[ht]
	\begin{center}
		\includegraphics[bb=260 20 1470 1065,width=7cm,height=6cm]{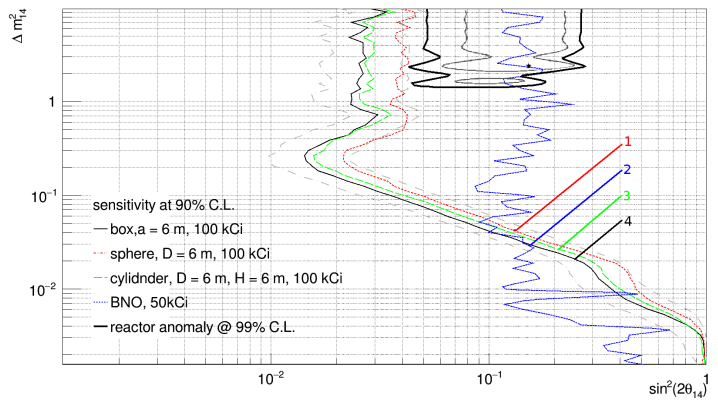}
		\caption{ Maximal theoretical sensitivity to sterile neutrino oscillation parameters at 90\% C.L.: 1 - spherical detector (90 t), 2 - Baksan neutrino observatory (330 t), 3 - cylindrical detector (135 t) , 4 - cubic detector (172 t).
			The bold lines and the star show reactor neutrino anomaly parameters region at 95\% and 99\% C.L. and the best-fit value (star). The dash-dotted lines illustrate the width of statistical jitter of a single experimental realization sensitivity curve within 1$\sigma$. 
			The region is covered only partially for the current BNO experiment configuration and currently available 50 kCi $^{144}\rm{Ce}$ source. 
			Increase of experimental statistics allows to fully cover reactor anomaly region at 95\% C.L. for cubic and cylindrical geometries that show similar performance. 
		}
		\label{Fig:sensitivity}
	\end{center}
\end{figure}

The statistical test was based on maximum likelihood approach with the binned Poisson likelihood in form of "effective $\chi^2$" computed on antineutrino spectrum S(E,R):
\[
   -log(L) =  \sum_{E_i,R_j}\frac{1}{2}\times(\lambda + log \Gamma(K+1)-log(\lambda)K),
\] 
 where $\lambda$ is the "theoretical" spectrum derived from Monte-Carlo simulation with statistics corresponding to 5 MCi by multiplication to neutrino survival probability for given $\Delta m_{14}$ and $sin^2(2\theta_{14})$ and scaled down by a statistical factor and K is the result of Monte-Carlo simulation with "experimental" statistics corresponding to 100 (50) kCi source.
 The sum is performed over the binned antineutrino spectrum $S(E_i,R_j)$.
 The survival probability is taken as two-neutrino approximation as the sterile eigenstate is expected to be much heavier than the known neutrinos \cite{Strumia:2006db}:
 \[
     F(E,R) = 1-sin^2(2\theta_{14})sin^2(1.27\Delta m_{14}^2[eV]\frac{R[m]}{E[MeV]})
 \]  
 where $\Delta m_{14}^2$ is the difference  of mass squares for sterile and active neutrinos, $\theta_{14}$ is the mixing angle, R is the distance between the source and registration point and E is the antineutrino kinetic energy.
 The sensitivity limits were obtained from the likelihood function profile built for various values of $\Delta m_{14}^2$ and mixing. The operation was performed for 100 realizations of test statistics, that were used for understanding of possible statistical variations of the experiment sensitivity. 
 
 The results of the study are shown on fig.~\ref{Fig:sensitivity}. The figure shows that the BNO detector has the best sensitivity to low values of $\Delta m_{14}^2$, while the cubic and cylindrical geometries show practically equal sensitivity exceeding results of the spherical detector geometry. Since the cylindrical geometry has the advantage of the lowest mass and easier constructive availability, it seems to be the optimal option among the ones tested.
 
\begin{figure}[ht]
 	\begin{center}
 		\includegraphics[bb=260 20 1470 1065,width=6cm,height=7cm]{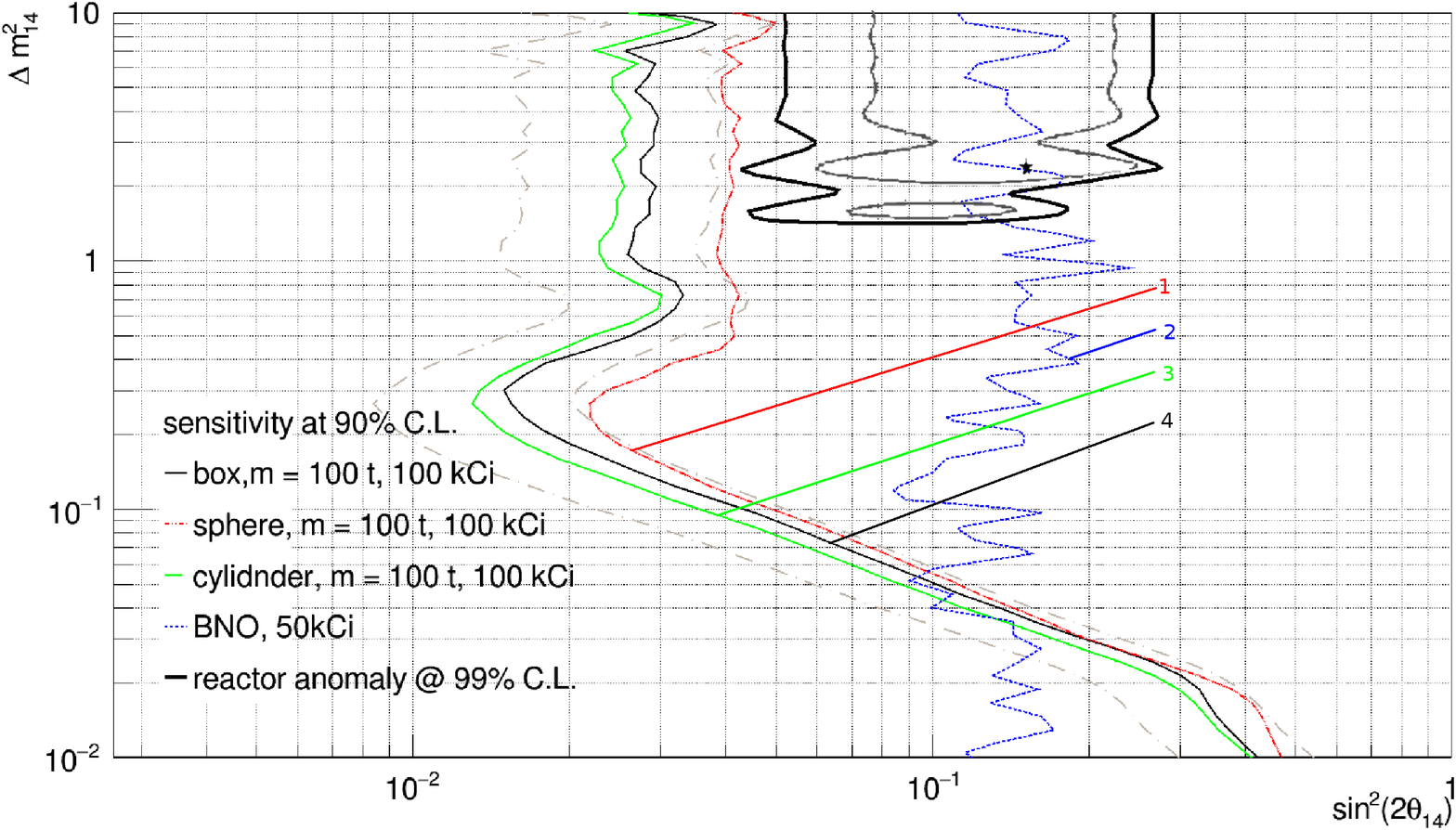}
 		\caption{ The same as in fig.~\ref{Fig:sensitivity}, but the detectors in the form of a sphere, cylinder and cube have the same mass of 100 tons: 1 - spherical detector (100 t),  2 - Baksan neutrino observatory (330 t), 3 - cylindrical detector (100 t) , 4 - cubic detector (100 t).
 		}
 		\label{Fig:sensitivity_100t}
 	\end{center}
 \end{figure}

 Since many parameters and characteristics of the future experiment are determined by the full mass of the scintillator, we checked the sensitivity of the detectors of the three above-mentioned shapes with the same mass of the scintillator. Such a comparison is shown in fig.~\ref{Fig:sensitivity_100t}. Again, the cylindrical shape of the detector provides better sensitivity to oscillation parameters.
 
It appears that the activity and geometrical limits we considered  allow to cover the reactor antineutrino anomaly region. The same time the BNO detector with 50 kCi source does not have enough sensitivity to perform the test of this anomaly. 

Full coverage of this region at 90 \% C.L. could be achieved with the considered activity and detector mass (e.g. to 100 t and 100 kCi) and solid angle with respect to the source in case of availability of zero-background regime.  
One should note that increase of the detector linear size influences sensitivity on $\Delta m_{14}$, while sensitivity on mixing comes mostly from the experiment statistics, limited by the detector volume and source activity.  
The optimal setup parameters will be refined in further investigations. The further investigations will also include consideration of the detector backgrounds coming from natural radioactivity and the $^{144}$Ce - $^{144}$Pr source. 
 
\section{Conclusions}
A sensitivity study for sterile neutrino oscillation parameters with various detector geometries has been performed. The sensitivity to sterile neutrino mixing angle for neutrino mass above 0.2 eV is around $sin^2(2\theta_{14}) < 0.03$. 
It was found out that the linear detector sizes of 6 m combined with 100 kCi source provide enough sensitivity to fully cover reactor neutrino anomaly region. The optimal detector geometry is chosen to be cylindrical. The optimal parameters of the setup, e.g. length to diameter ratio, could be refined with usage of the software developed. This will be done in the further investigations.

This means that satisfactory sensitivity to neutrino oscillation to sterile eigenstate with 1 eV mass could be achieved only in case the statistics studied, namely with 100 kCi $^{144}$Ce - $^{144}$Pr source and detector with linear dimensions of 6 m.

 This work was supported by RSF, project 17-12-01009 and RFBR, project 16-29-13014.

\bibliographystyle{unsrtnat}

\end{document}